\documentclass[aps, preprint, showpacs, showkeys, amsmath, floatfix]{revtex4}

\usepackage[dvipdfmx]{epsfig}
\usepackage{color}

\usepackage{ulem}
\usepackage{amsmath}	% required for `\cases' (yatex added)

\begin{document}

\title{Stochastic theory of polarized light in nonlinear birefringent media: An application to 
optical rotation}

\date{\today}

\author{Satoshi Tsuchida}
\email[E-mail: ]{stsuchida88@gmail.com}
\affiliation{Department of Physics, Osaka City University, 3--3--138 Sugimoto, Sumiyoshi--ku, Osaka City, Osaka, 558--8585, Japan}
\author{Hiroshi Kuratsuji}
\affiliation{Department of Physics, Ritsumeikan University-BKC, 1--1--1 Noji--higashi, Kusatsu City, Shiga, 525--8577, Japan}

\begin{abstract}

A stochastic theory is developed for the light
transmitting the optical media exhibiting linear and nonlinear birefringence.
The starting point is the two--component nonlinear Schr{\"o}dinger equation (NLSE).
On the basis of the ansatz of ``soliton'' solution for the NLSE,
the evolution equation for the Stokes parameters is derived,
which turns out to be the Langevin equation by taking account of randomness
and dissipation inherent in the birefringent media.
The Langevin equation is converted to the Fokker--Planck (FP) equation for
the probability distribution by employing the technique of functional integral
on the assumption of the Gaussian white noise for the random fluctuation.
The specific application is considered for the optical rotation,
which is described by the ellipticity (third component of the Stokes parameters) alone:
(i) The asymptotic analysis is given for the functional integral,
which leads to the transition rate on the Poincar{\'e} sphere.
(ii) The FP equation is analyzed in the strong coupling approximation,
by which the diffusive behavior is obtained for the linear and nonlinear birefringence.
These would provide with a basis of statistical analysis for
the polarization phenomena in nonlinear birefringent media. 
\end{abstract}


\maketitle

\section{Introduction}
\label{sec:intro}

The wave propagation in random media has been studied as a specific problem (e.g.~\cite{Dashen,Shapiro,Segev,Ishimaru})
in the wider discipline of the random phenomena ranging from molecular level to cosmological scale
\cite{Chandra,Uhlen,Kadanoff,Kubo,Dyson}.
The randomness is caused by the presence of various sorts of fluctuations inherent in external agents.
Turning to quantum physics, the stochastic approach to Schr{\"o}dinger equation has been worked out; e.g.~in Refs.~\cite{Someda,Biele},
which employ the stochastic differential equation to describe the several ways of random fluctuations inherent in quantum waves.
We also mention the work concerning physics of random potential~\cite{Langer,Helfand}.
In our previous work the stochastic analysis was investigated for the generalized Schr{\"o}dinger
equation in the presence of the random fluctuations~\cite{Tsuchida}.

Now one often encounters the Schr{\"o}dinger type equation in physics; typically,
the optical (or electromagnetic (EM)) waves that propagate through linear and/or nonlinear medium.
This is called the nonlinear Schr{\"o}dinger equation (NLSE) and  has been investigated for long time 
in wave guide formalism
\cite{Chiao,Snyder,Swartz,Christou}.
The EM (or optical) wave is also characterized by the polarization degree of freedom,
which plays a crucial role when the light transmits optically anisotropic medium~\cite{LL,BW}.
So it is intriguing to explore a problem of the polarized light in the stochastic framework.

In this paper we develop a stochastic theory for the polarized light in nonlinear birefringent media.
The starting point is the two component NLSE~\cite{Kra} which is arranged to take account of the polarization degrees of freedom (PDOF).
Here the crux is that the PDOF is extracted from {\it{nonlinear birefringence}}~\cite{Maker}
by using the {\it collective coordinate method} used in the nonlinear field theory model~\cite{Skyrme}
(see also the works on the random as well as thermal effect on the soliton motion occurring in 
the Sine-Gordon model\cite{Marchesoli,Hanggi}) . 
The resultant equation is given in terms of the Stokes parameters~\cite{Sala,Tratnik,KK,Brosseau},
which turns out to be the Langevin equation in the presence of the randomness together with the effect of dissipation.
The Langevin equation is further converted to the Fokker--Planck (FP) equation by using the functional integral
technique on the basis of the assumption of the Gaussian white noise for the random fluctuations.

After presenting the general stochastic formulation, we apply this
to a specific class of birefringent media characterized by the {\it optical rotation},
which is designed to describe the linear as well as nonlinear birefringence.
Specifically we scrutinize the stochastic behavior of {\it{ellipticity}}
accompanying the Faraday effect and/or nonlinear birefringence.
The procedure is carried out by two aspects:
(i) One is an asymptotic analysis of the functional integral,
which enables us to evaluate the transition amplitude in the framework of the asymptotic limit
by using an analogy with the semiclassical approximation of quantum mechanics.
(ii) The other is an analysis of the FP equation,
which serves as a way to explore the diffusive behavior of the distribution function on the
Poincar{\'e} sphere leading to the equilibrium state described by Boltzmann distribution
as a consequence of the fluctuation-dissipation relation.

The paper is organized as follows:
In the sections from~\ref{sec:preliminary} to~\ref{sec:func_int_fp}, the general formalism is constructed.
The brief account has been presented in the previous note~\cite{Tsuchida2}.
In~\ref{sec:fieldeq} a preliminary account is given of the two component NLSE.
In section~\ref{sec:evolution_sp}, the equation of motion of the Stokes parameters
by using the soliton wave for the single component NLSE.
In section~\ref{sec:le_fi} the Langevin equation is derived by incorporating the random fluctuation and the dissipation.
In section~\ref{sec:func_int_fp}, the functional integral is constructed on the basis of the ansatz of the Gaussian white noise,
by which the FP (Smolkowski) equation is derived.
Section~\ref{sec:application} is the application to the optical rotation is analyzed
on the basis of the general formation developed in section~\ref{sec:func_int_fp}.

\section{Preliminary}
\label{sec:preliminary}

\subsection{Field equation of the polarized light}
\label{sec:fieldeq}

The standard procedure to explore the evolution of light is based on
{\it para-axial scheme} (assisted by an envelope approximation) \cite{Yariv},
which describes the light propagation along a prescribed direction that is perpendicular to the polarization plane.
The starting point is the two--component wave equation for the electric field $ {\bf{E}} $:
\begin{eqnarray}
  \label{eq:fieldeqE}
  \frac{ {\partial}^2 {\bf{E}} }{ \partial z^2 } + \left( \frac{ {\partial}^2 }{ \partial x^2 } + \frac{ {\partial}^2 }{ \partial y^2 } \right) {\bf{E}} + \hat{\epsilon} {\bf{E}} = 0
\end{eqnarray}
Here $ {\bf{E}} $ represents two--component {\it{complex}} wave function,
which is in the $ ( x, y ) $ plane orthogonal to the propagation direction.
We adopt the following setting of the coordinate:
$ z $ stands for the coordinate of propagation direction,
and $ {\boldsymbol{x}} = ( x, y ) $ is the coordinate of polarized plane.
The anisotropic nature of the optical media is incorporated in the dielectric tensor $ \hat{\epsilon} $,
which includes both linear and nonlinear birefringence.
We shall reduce Eq.~(\ref{eq:fieldeqE}) to the two--component NLSE~\cite{Chiao,Swartz}. 
We write the electric field in the following form:
\begin{eqnarray}
  \label{eq:elefield}
  {\bf{E}} ({\boldsymbol{x}}, z) = {\bf{f}} ({\boldsymbol{x}}, z) e^{i k n_{0} z},
\end{eqnarray}
which represents the plane wave with modified amplitude $ {\bf{f}} ( {\boldsymbol{x}}, z ) $.
$ k = {\omega \over c} $ is the wave number, and $ n_{0} ( \equiv \sqrt{\varepsilon_{0}} ) $
means the refractive index for the case as if the medium is isotropic.
The amplitude $ {\bf{f}} ( {\boldsymbol{x}}, z ) $ is written as 
$ {\bf{f}} = {}^t(f_{1}, f_2) = f_1 {\bf{e}}_{1} + f_2 {\bf{e}}_{2} $,
where $ {\bf{e}}_{1} $ and $ {\bf{e}}_{2} $ denote the basis of linear polarization.
Here we adopt the {\it envelope approximation}~\cite{Yariv,KK,Kra}:
Namely it is assumed that $ {\bf{f}} $ is slowly varying function of $ z $, that is
$ \left| \frac{ {\partial}^{2} {\bf{f}} }{ {\partial} {z}^{2} } \right| \ll k \left| \frac{ {\partial} {\bf{f}} }{ \partial z } \right| $.
Under this assumption, the field equation~(\ref{eq:fieldeqE}) can be reduced to the wave equation
for the modified amplitude $ {\bf{f}} $:
\begin{equation}
  i\lambda\frac{\partial {\bf f}}{\partial z} + \frac{ {\lambda}^{2} }{ 2 n_{0} }
  \nabla^2_{\bot}{\bf f} - \hat v {\bf f} = 0 
\end{equation}
where $ \lambda $ stands for the wave length devided by $ 2 \pi $,
$ {\nabla}_{\bot}^{2} \equiv \frac{ {\partial}^{2} }{ {\partial} x^{2} } + \frac{ {\partial}^{2} }{ {\partial} y^{2} } $
means the Laplacian with respect to the polarized plane,
and $ \hat v = - \frac{1}{2 n_{0}} \left( {\lambda}^{2} \hat\epsilon - n_0^2 \right) $
represents the potential written in terms of $ 2 \times 2 $ matrix. 

In order to describe the evolution of the polarized state in the following argument,
it is convenient to transform the basis to the circular basis instead of the linear polarization 
($ {\bf{e}}_1, {\bf{e}}_2 $), which is written as 
\begin{eqnarray}
  \left(
    \begin{array}{c}
      e_{+} \\
      e_{-}
    \end{array}
  \right)
  = {1 \over \sqrt{2}} \left(
    \begin{array}{cc}
      1 & i \\
      1 & -i 
    \end{array}
\right)
  \left(
    \begin{array}{c}
      e_{1} \\
      e_{2}
    \end{array}
  \right)
  \equiv T
  \left(
    \begin{array}{c}
      e_{1} \\
      e_{2}
    \end{array}
  \right)
\end{eqnarray}

The corresponding transformation for the two component wave is given by 
\begin{equation}
  \psi = U{\bf f}= {}^t\left( \psi_1, \psi_2 \right)
\end{equation}
Here $ U $ is defined as the complex conjugate of $ T $; namely, $ U = T^{*} $.
Thus we have the two component Schr{\"o}dinger-type equation for $ \psi $:
\begin{equation}
  \label{eq:NLSE}
  i\lambda{\partial \psi \over \partial z} = {\cal{H}} \psi
\end{equation}
with the transformed ``Hamiltonian''
\begin{eqnarray}
  {\cal{H}} = - \frac{ {\lambda}^{2} }{ 2 n_{0} } {\nabla}_{\bot}^{2} + V ~~~,~~~ V = U \hat{v} U^{-1} = V_{\rm{NL}} + V_{\rm{L}} 
\end{eqnarray} 
Note that $ z $ plays a role of the {\it time} variable.
Equation~(\ref{eq:NLSE}) is a typical para--axial equation that has been widely known in the optical physics~\cite{LL,BW}.
The explicit form of $V_{\rm{L}}$ and $V_{\rm{NL}}$ are given as follows:
(We put $ g $ in the second term of $ V_{\rm{NL}} $ as positive value.)
\begin{eqnarray}
  V_{\rm{L}} = \frac{G}{2} \left( e_{1}^{2} + e_{2}^{2} \right) {\bf{1}} +
  \left(
    \begin{array}{cc}
      {\gamma} & {\alpha} - i {\beta} \\
      {\alpha} + i {\beta} & - {\gamma}
    \end{array} 
  \right)
  \label{eq:vl} \\
  V_{\rm{NL}} = \frac{ G_{0} }{ 2 } \vert {\psi} \vert^{2} {\bf{1}} - g
  \left(
    \begin{array}{cc}
      - \frac{ \vert {\psi}_{1} \vert^{2} - \vert {\psi}_{2} \vert^{2} }{ 2 } & {\psi}_{2}^{\dagger} {\psi}_{1} \\
      {\psi}_{1}^{\dagger} {\psi}_{2} & \frac{ \vert {\psi}_{1} \vert^{2} - \vert {\psi}_{2} \vert^{2} }{ 2 }
    \end{array} 
  \right)
  \label{eq:vnl} \nonumber
\end{eqnarray}
From these equations, one can see that
the potential $ V_{\rm{L}} $ and $ V_{\rm{NL}} $ serve as the contribution coming from the linear and nonlinear birefringence, respectively.
That is, $ V_{\rm{L}} $ does not depend on the wave field $ \psi $, whereas the $ V_{\rm{NL}} $ depends on $ \psi $.
The parameters $ {\alpha} $, $ {\beta} $ and $ {\gamma} $ in $ V_{\rm{L}} $ are described by
the effects of external electric and magnetic field, say, $ {\bf E}_{ext}= (e_{1}, e_{2}) $ and $ {\rm{H}}_{\rm{ex}} $.
$G$ means the coupling strength for the external Kerr effect (see, appendix~\ref{append:derivation}).

It is to be noted that the  parameters $ (\alpha, \beta, \gamma) $ possess with the property that could be
controlled by external conditions in various ways: namely, it is apparently possible to control these by
modulating the external electric and magnetic fields.  So one expects a possible effect caused by
such modulation, which means that the parameters are allowed to be time-dependent in general.

On the other hand, the first term in $ V_{\rm{NL}} $ gives the {\it scalar type} nonlinear interaction with the coupling constant $ G_{0} $,
which gives rise to a solution that is used to construct a trial form of the polarized wave (see next section).
The second term just represents the nonlinear birefringence that is characterized by the coupling constant $ g $ \cite{Maker,Sala},
which plays a key role to govern the evolution of the Stokes parameters. These coupling constant should not be controlled externally, of course.

\subsection{The evolution of the Stokes parameters}
\label{sec:evolution_sp}

\subsubsection{Equation of motion inspired by a soliton solution}
\label{subsec:eom_ss}

Now we try to extract PDOF by looking for a specific solution of the two--component NLSE.
In order to carry out this, we adopt the following procedure;
let us first consider the ``soliton'' type solution
(which is familiar in nonlinear optics~\cite{Chiao, Snyder, Swartz, Christou})
of the single component NLSE:
\begin{eqnarray}
  \left[ - \frac{ {\lambda}^{2} }{ 2 n_{0} } {\nabla}_{\bot}^{2} + \frac{ G_{0} }{ 2 } \vert F \vert^{2} + k \right] F({\boldsymbol{x}}) = 0 \nonumber
  \label{eq:solitonsolu}
\end{eqnarray}
where $ k = \frac{G}{2} \left( e_{1}^{2} + e_{2}^{2} \right) $ is the first term of $ V_{L} $.
By utilizing the solution $ F \left( {\boldsymbol{x}} \right) $,
we put an ansatz for $ \psi $ by using the spinor form:
\begin{eqnarray}
  {\bf{\psi}} = F \left( {\boldsymbol{x}} \right)
  \left(
    \begin{array}{c}
      a_1 (z) \\
      a_2 (z)
    \end{array}
  \right)
  \equiv F \left( {\boldsymbol{x}} \right) \tilde \psi 
\end{eqnarray}
Here we write the spinor in terms of the angular variable:
\begin{eqnarray}
  \tilde \psi = 
  \left(
    \begin{array}{c}
      \cos \frac{\theta (z) }{2} \\
      \sin \frac{\theta (z) }{2} e^{ i {\phi} (z) }
    \end{array}
  \right)
  \label{eq:ansatzpsi}
\end{eqnarray}

In order to describe a dynamical behavior of the angular variable $ (\theta, \phi) $, we use the action function
$$  I = \int {\cal{L}} d^{2} x dz 
$$ 
with $ {\cal{L}} $ being the Lagrangian density, which is written as
\begin{eqnarray}
  I  = \int {\psi}^{\dagger} \left( i {\lambda} \frac{{\partial}}{{\partial}z} - {\cal{H}} \right) {\psi} d^{2} x dz  
  \label{eq:actionlag}
\end{eqnarray}
Here the effective Lagrangian $ {\cal{L}}_{eff} $ is defined for the PDOF, which is written as the sum:
\begin{equation}
  \mathcal{L}_{eff} = \frac{1}{A} \int {\cal{L}} d^{2} x = {\cal{L}}_{1} - H
\end{equation}
with $ H $ being the Hamiltonian; $ H = \frac{1}{A} \int \psi^{\dagger} {\cal{H}} \psi d^{2} x $,
and $ A = \frac{\lambda}{2} \int F^{2} ({\bf{x}}) d^{2} x $.
$ \mathcal{L}_1 $ and $ H $ are given in terms of the angular variables $ (\theta, \phi) $:
\begin{eqnarray}
  \mathcal{L}_1  & = &  - \left( 1 - \cos \theta(z) \right) \frac{ d {\phi} (z) }{ dz } \nonumber \\
  H & = & \frac{2}{\lambda} \left[ {\alpha} \sin {\theta}(z) \cos {\phi}(z) + \beta \sin {\theta}(z) \sin {\phi}(z) + \gamma \cos {\theta}(z) \right] \nonumber \\
  & & + \frac{1}{2} B \cos 2{\theta}(z) - C
\end{eqnarray}
where the parameters $ B $ and $ C $ are given as
\begin{eqnarray}
  B &=& \frac{ g }{ A } \int F^{4} ({\boldsymbol{x}}) d^{2} x \\
  C &=& \frac{ 1 }{ A } \int \left[ \frac{ {\lambda}^{2} }{ 2 n_{0} } F({\boldsymbol{x}}) {\nabla}_{\bot}^{2} F({\boldsymbol{x}}) 
    - \frac{ G_{0} }{ 2 } F^{4} ({\boldsymbol{x}}) - k F^{2} ({\boldsymbol{x}}) \right] d^{2} x \nonumber
\end{eqnarray}
Thus the nonlinear feature of polarized light can be incorporated in these values ($A$, $B$ and $C$)
together with the specific values ($\alpha$, $\beta$ and $\gamma$) of the linear birefringence.

We now introduce the Stokes parameters as follows:
\begin{eqnarray}
  {\bf{S}}_{i} = \tilde{{\bf{\psi}}}^{\dagger} {\sigma}_{i} \tilde{{\bf{\psi}}} ~~,~~
  {\bf{S}} = \left( \sin {\theta} \cos {\phi}, \sin {\theta} \sin {\phi}, \cos {\theta} \right), 
  \label{eq:stokespara}
\end{eqnarray}
which represents the point of the Poincar{\'e} sphere,
and is designated by the angular coordinates $ \left( \theta, \phi \right) $,
that is, the polarized state is completely described by the point on the sphere in geometrical way.
The general point on the Poincar{\'e} sphere represents the elliptic polarization and
$ \cos \theta $ serves as an {\it ellipticity} of the light polarization.
In particular, $ \theta = 0 $ and $ \theta = \pi $ represent
left--handed and right--handed circular polarization, respectively,
and $ \theta = \frac{ \pi }{ 2 } $ means the linear polarization (see Fig.1).  
%~\ref{fig:poincare}).

%
\begin{figure}[t]
  \begin{center}
    \includegraphics[width=60mm]{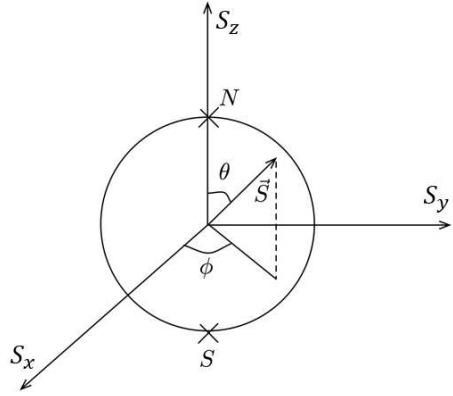}
    \caption{(Color Online).
     Poincar{\'e} sphere.}
   \label{fig:poicare}
  \end{center}
\end{figure}

The Hamiltonian $ H $ is written in terms of the Stokes parameters:
\begin{eqnarray}
  H  = \frac{ 2 }{ \lambda } \left[ {\alpha} S_{x} + {\beta} S_{y} + {\gamma} S_{z} \right] + \frac{1}{2} B \left( 2 S_{z}^{2} - 1 \right) - C
  \label{eq:stokeshamil}
\end{eqnarray}
By using the variation principle $  \delta \int \mathcal {L}_{eff} dz = 0 $,
we can obtain the coupled equation for $ \theta (z) $ and $ \phi (z) $, which results in 
\begin{equation}
 \dot\theta = \frac{1}{\sin\theta}\frac{\partial H}{\partial \phi},  
~~\dot\phi = - \frac{1}{\sin\theta}\frac{\partial H}{\partial \theta}. 
\end{equation}
The equation of motion for the Stokes parameters $ {\bf S} $ is slightly complicated, which is derived from the variational (action) principle  
\begin{equation}
  \delta I =  \int (\delta\omega - \delta H dz ) = 0 
 \label{action}
\end{equation}
Here  it is crucial to rewrite this in terms of the variation of the Stokes parameters for introducing 
the dissipative effect in what follows, namely we write the term $ \delta\omega $ 
\begin{eqnarray}
  \delta \omega & = & \left[ \left( S_{x} \frac{dS_{y}}{dz} 
      - S_{y} \frac{dS_{x}}{dz} \right) \delta S_{z} + \left( S_{y} \frac{dS_{z}}{dz} - S_{z} \frac{dS_{y}}{dz} \right) \delta S_{x} 
    + \left( S_{z} \frac{dS_{x}}{dz} - S_{x} \frac{dS_{z}}{dz} \right) \delta S_{y} \right] dz \nonumber \\
  & = & \left[ \left({\bf{S}} \times \frac{d{\bf{S}}}{dz} \right) \cdot \delta {\bf{S}} \right] dz  \nonumber 
\end{eqnarray}
together with $ \delta H dz = \left[\frac{\partial H}{\partial {\bf{S}}} \cdot \delta {\bf{S}} \right] dz $.  Hence 
we have the variational equation: 
\begin{eqnarray}
  \label{eq:eom_principle}
  \frac{ \delta I }{ \delta {\bf{S}} } = 0
\end{eqnarray}
which leads to the equation of motion 
\begin{eqnarray}
  {\bf S} \times \frac{ d {\bf{S}} }{ dz } = \frac{ {\partial} H }{ {\partial} {\bf{S}} }
  \label{eq:eom_stokes1}
\end{eqnarray}
or it can be rewritten in the form 
\begin{eqnarray}
\frac{ d {\bf{S}} }{ dz } = - {\bf{S}} \times \frac{ {\partial} H }{ {\partial} {\bf{S}} }
  \label{eq:stokes1}
\end{eqnarray}
This corresponds to the equation of motion for a real spin.
More precisely, this may be regarded as the equation of motion for the spin of light field,
since the concept of Stokes parameters is a classical counterpart of the photon spin.

\subsection{Specific birefringence: Optical rotation}

We shall examine the specific solutions for the equation of motion of the Stokes parameters, i.e. Eq. (\ref{eq:stokes3}),
which enable one to obtain a pictorial view of the polarization.
Here we restrict the argument to specific birefringence $ {\alpha} = {\beta} = 0 $.
Further we take two cases
(i) $ \gamma \neq 0,~B = 0 $,
and (ii) $ \gamma \neq 0, B \neq 0 $,
namely, the pure Faraday effect without and with the nonlinear birefringence. 

For the case (i), $ H = \frac{2}{\lambda} \gamma \cos\theta $,
the equation of motion simply becomes 
$ \dot\theta = 0 ~,~\dot\phi = \frac{2}{\lambda} \gamma $ 
which leads to $ \theta = \theta_0 ,~ \phi = \frac{2}{\lambda} \gamma z + \phi_0 $.
Alternatively, this is written in terms of the Stokes parameters: 
\begin{eqnarray}
  S_{x} = \sin {\theta}_{0} \cos \left( \frac{2}{\lambda} {\gamma} z + {\phi}_{0} \right) ~~,
  ~~ S_{y} = \sin {\theta}_{0} \sin \left( \frac{2}{\lambda} {\gamma} z + {\phi}_{0} \right) ~~,
  ~~ S_{z} = \cos {\theta}_{0}
  \label{eq:solutionstokes}
\end{eqnarray}
which traces a circle with constant latitude $ \theta_0 $. 

For the case (ii), $ \gamma \neq 0 ~,~ B \neq 0 $, we can simply derive the equation of motion;
\begin{eqnarray}
  \label{eq:solucase2}
  \dot{\theta} = 0 ~~,~~ \dot{\phi} = \frac{2}{\lambda} \gamma + 2 B \cos \theta
\end{eqnarray}
which gives the solution by using the Stokes parameters:
\begin{eqnarray}
  \label{eq:solutionstokes}
  S_{x} = \sin {\theta}_{0} \cos \left( Dz + {\phi}_{0} \right) ~~,~~ 
  S_{y} = \sin {\theta}_{0} \sin \left( Dz + {\phi}_{0} \right) ~~,~~ S_{z} = \cos {\theta}_{0}
\end{eqnarray}
Here $ D = \frac{2}{\lambda} {\gamma} - 2B S_{z} $,
which includes the modification by the nonlinear birefringence.
This also represents the uniform rotation of the elliptic polarization
characterized by the constant latitude $ \theta_{0} $.
This can be regarded as a generalization of the Faraday effect.
From the arguments just above, both cases are called the {\it optical rotation}.

{\it{Remark}}:
Here we give a brief comment on the other type of linear birefringence which differs from the optical rotation.
The birefringence is given by arranging such that
$ {\alpha} \neq 0 ~,~ {\beta} = {\gamma} = 0 ~,~ B \neq 0 $,
which describes the simultaneous presence of the external Kerr effect and the nonlinear birefringence.
This case is much more complicated: The solution for Stokes parameters is given in terms of
the elliptic functions and the brief explanation will be presented in the appendix,
so in what follows we do not discuss it any more.

\section{Langevin equation: The inclusion of dissipation and random fluctuations} 
\label{sec:le_fi}

Now we study the random effect on the polarized light.
The Stokes parameters are particularly useful to describe the effect,
because it has the similar structure with the magnetic spin~\cite{Brown}.
The effect of fluctuations arises from various origins.
The most important factor may be the scattering of light (see e.g.~\cite{Tiggelen}).
This could be caused by impurities (e.g. micro particles),
which possess the property that gives rise to ``optical activity'' in a broad sense.
If these impurities are collocated as random configuration, the PDOF will acquire randomness by the scattering.
Then, it is natural to expect that the Brownian motion of the Stokes parameters occurs as the result of these processes.
In what follows we have a mind of the case that the fluctuation effect gives rise to the
randomness $ {\bf{\xi}} $ for the ``pseudo-magnetic field'', so we can replace
$ \frac{\partial H}{\partial {\bf{S}}} \rightarrow \frac{\partial H}{\partial {\bf{S}}} + {\bf{\xi}} $.
Here $ {\bf{\xi}} (z) $ is assumed to obey the Gaussian white noise.

In actual optical media, besides the random noise it is inevitable to take account of the effect of
the dissipation of the strength of light, that is caused by absorptions~\cite{LL, BW}.
One of such absorptions may be due to the transfer of the energy to the background radiation.
In the birefringent media, which is described by the Stokes parameters,
one can consider a peculiar feature coming from the similarity between the real spin and the Stokes parameters~\cite{Brown}.
Following the well known procedure in non--equilibrium statistical physics~\cite{LL2},
the dissipative effect can be given in a phenomenological form;
$ \frac{\partial H}{\partial {\bf{S}}} \rightarrow \frac{\partial H}{\partial {\bf{S}}} + \mu \frac{d{\bf{S}}}{dz} $.
This replacement is justified as follows:
The right hand side of Eq.~(\ref{eq:eom_principle})
is modified by adding the term that is derived from the dissipative function
$ K = \frac{ \mu }{ 2 } \left( \frac{ d {\bf{S}} }{ dz } \right)^2 $,
namely, it is replaced by
\begin{equation}
  \frac{\delta I}{\delta {\bf S}} ( = \frac{\partial K}{\partial \dot{\bf S}} ) =  \mu\frac{d{\bf S}}{dz} 
\end{equation}
Thus by taking account of the randomness as well as the dissipation the equation of motion is modified as
\begin{eqnarray}
  \label{eq:stokes2}
  {\bf{S}} \times \frac{ d {\bf{S}} }{ dz } =  \frac{ {\partial} H }{ {\partial} {\bf{S}} } + \mu  \frac{ d {\bf{S}} }{ dz } + \xi 
\end{eqnarray}
We also have another equation: 
\begin{equation}
  \label{eq:stokes3}
  \frac{ d {\bf{S}} }{ dz } = -{\bf S} \times  \left( \frac{ {\partial} H }{ {\partial} {\bf{S}} } + \mu  \frac{ d {\bf{S}} }{ dz } + \xi \right) 
\end{equation}
From Eqs. (\ref{eq:stokes2}) and (\ref{eq:stokes3}),
$ \frac{d{\bf S}}{dz} $ is solved and hence we have the Langevin equation
\begin{eqnarray}
  \frac{d{\bf S}}{dz} = - {\bf{A}} + {\bf{\eta}}
  \label{eq:langevinnoise}
\end{eqnarray}
Here $ {\bf A} $ and $ \eta $ are given as
\begin{equation}
  {\bf{A}} =  \frac{ 1 }{ 1 + {\mu}^{2} } \left[ {\mu} \frac{ {\partial} H }{ {\partial} {\bf{S}} } 
    + {\bf{S}} \times \frac{ {\partial} H }{ {\partial} {\bf{S}} } \right],  
  {\bf{\eta}} = - \frac{ 1 }{ 1+ {\mu}^{2} } \left[ \mu \xi + {\bf{S}} \times {\bf{\xi}} \right]  
  \label{eq:landaulif}
\end{equation}
$ \eta $ just gives the resultant random noise.
Without $ \eta $, the Langevin equation~(\ref{eq:langevinnoise}) is
an analogy of the ``Landau-Lifschitz equation'',
that is well known the ferromagnetic theory~\cite{LG}.
Note that according to this equation of motion, the magnitude of the light field is conserved:
$ {\bf S}\cdot \frac{d{\bf S}}{dz} = \frac{d({\bf S})^2}{dz} = 0 $,
where we have used the relation $ {\bf S} \cdot \frac{\partial H}{\partial {\bf S}} = 0 $.
This feature is consistent with the behavior of the orbit on the Poincar{\'e} sphere as will be seen below.
On the contrary, the energy is dissipative, as is verified by the relation:
$ \frac{dH}{dz}  = - \frac{ \mu }{1 + \mu^2} (\nabla H)^2 \leq 0 $.

\section{Functional integral and the Fokker-Planck equation}
\label{sec:func_int_fp}

From the Langevin equation (\ref{eq:langevinnoise}),
we can derive the FP equation for the probability distribution of Stokes parameters.
This is carried out by employing the method of functional integral.

\subsection{Functional integral}
\label{general}

{\it Structure of the Langevin equation}: 
Before deriving functional integral, we need to characterize the nature of random noise.
First to be mentioned is that we are concerned with the stochastic process on the Poincar{\'e} sphere.
In order to reveal this feature, it is convenient to use the spherical coordinate instead of the rectangular coordinate.
If we consider the situation that the noise $ {\xi} $ and $ \eta $ lie on the Poincar{\'e} Sphere,
these direct to the tangential space, namely $ {\bf{S}} \cdot \xi = {\bf{S}} \cdot \eta = 0 $.
This way, by using a spherical basis ($ \hat r, ~\hat \theta, ~\hat \phi $),
we can write $ {\bf{S}} = \hat{r} $ and $ \xi = {\xi_{\theta}} \hat{\theta} + \xi_{\phi} \hat{\phi} $.
Thus, we obtain $ {\bf{S}} \times \xi = - \xi_{\phi} \hat{\theta} + \xi_{\theta} \hat{\phi} $.
Hence, we have the transformation between $ \xi $ and $ \eta $ such that
\begin{eqnarray}
  \eta_{\theta} = - \frac{ 1 }{ 1 + {\mu}^{2} } \left( {\mu} {\xi}_{\theta} - {\xi}_{\phi} \right) \ \ , \ \
  \eta_{\phi} = - \frac{ 1 }{ 1 + {\mu}^{2} } \left( {\mu} {\xi}_{\phi} + {\xi}_{\theta} \right)
  \label{eq:trans_xi_eta}
\end{eqnarray}
This is essentially orthogonal transformation except for a scale factor,
which means that the resultant noise $ \eta $ is simply obtained by rotating the original noise $ \xi $.
Note that the norm is preserved under orthogonal transformation:
$ \eta^2 = \eta_{\theta}^{2} + \eta_{\phi}^{2} = \frac{ 1 }{ 1 + {\mu}^{2} } \xi^2 $.
In terms of the spherical basis, the Langevin equation is written as
\begin{eqnarray}
  \frac{ d \theta }{ dz } = - A_{\theta} + {\eta}_{\theta} \ \ , \ \ \sin \theta \frac{ d \phi }{ dz } = - A_{\phi} + {\eta}_{\phi}
\end{eqnarray}
where $ A_{\theta} $ ($ A_{\phi} $) means $ \theta $ ($ \phi $) component of the vector $ {\bf{A}} $ in Eq.~(\ref{eq:landaulif}).

At this point we can naturally introduce the Gaussian white noise for $ \eta = (\eta_{\theta}, \eta_{\phi}) $
which is characterized by the correlation:
\begin{eqnarray}
  \langle \eta_i(z)\eta_{j}(z+u) \rangle = \frac{ \kappa }{ 1 + {\mu}^{2} } \delta_{ij} \delta(u) \equiv 2h \delta_{ij} \delta(u)
  \label{GWN}
\end{eqnarray}
where the suffix $ (i, j) $ represents $ ( \theta, \phi ) $, $ \delta(u) $ means the delta function,
and $ h $ the diffusion constant (which corresponds to the Planck constant in quantum mechanics).
It should be noted that this is derived from the original noise $ \xi $, which is assumed to follow the Gaussian white noise:
$ \langle \xi_i(z)\xi_j(z+u) \rangle = \kappa \delta_{ij} \delta(u) $.
By applying the transformation Eq.~(\ref{eq:trans_xi_eta}) we obtain the Gaussian white noise $ \eta $.
In this way, by using the spherical basis, we have settled a physical meaning for the Langevin equation as the 
stochastic theory on the Poincare sphere.  

Now, following the above ansatz (\ref{GWN}),
the probability distribution is given by the standard Gaussian functional form: 
\begin{equation}
  P[{\bf{\eta}}(z)] = \exp \left[- \frac{1}{2h} \int_0^z  {\bf{\eta}}^2(z)dz \right] . 
\end{equation}
Using this distribution, the transition probability
from $ {\bf{S}} (0) $ to $ {\bf{S}} (z) $ is given by the functional integral
\begin{equation}
  K[{\bf S}(z)\vert {\bf S}(0)] = \int_{ {\bf{S}} (0) }^{ {\bf{S}} (z) }   \exp \left[- \int^{z}_{0} \frac{{\bf{\eta}}^2(z)}{2h}dz \right]
  \mathcal{D}[{\bf{\eta}}(z)]
\end{equation}
In order to obtain the functional integral for Stokes parameters ${\bf{S}}$,
we adopt the following steps: We insert the expression of the $\delta$--functional integral
\begin{equation}
  \int \delta[ {\bf F}(z) - \eta(z)] \mathcal{D}{\bf F}(z) = 1
\end{equation}
with $ {\bf F} = \frac{d{\bf S}}{dz} + {\bf{A}} $, and use the relation:
$ \delta[f(x)] \sim \int \exp[i\lambda  f(x)] d\lambda $.
After integrating over $ \eta $ and $ \lambda $, one gets
\begin{equation}
  K \sim \int \exp \left[ -\frac{1}{2h}\int_0^z  {\bf F}^2(z)dz \right] \mathcal{D}{\bf F} 
\end{equation}
Thus, we have the functional integral over the Stokes parameters, that is given as
\begin{equation}
  K[{\bf S}(z)\vert {\bf S}(0)] = \int  \exp 
  \left[- \frac{1}{2h}\int_0^z  \left(\frac{d{\bf S}}{dz} + {\bf A(S)} \right)^2dz \right] 
  J({\bf{S}}) \mathcal{D}[{\bf S}]. 
  \label{PI2}
\end{equation}
Here $ J({\bf{S}}) $ is the functional Jacobian, which is evaluated as follows~\cite{Tsuchida}:
\begin{equation}
  J({\bf{S}}) = {\rm{det}} \left( \frac{ \delta {\bf{F}}({\bf S} )}{\delta{{\bf S}}} \right) = \exp \left[ \int_{0}^{z} M dz \right]
\end{equation}
The factor $ M $ is given as 
\begin{equation}
  M = \frac{1}{2} \frac{ {\partial} }{ {\partial} {\bf{S}} } \cdot {\bf{A}}
  = \frac{1}{2} \frac{\mu}{1 + \mu^2} \nabla^2 H 
\end{equation}

\subsection{The Fokker-Planck (FP) equation on the Poincar{\'e} sphere}
\label{sec:fpeq}

Now we shall derive the FP equation for the distribution function on the Poincar{\'e} sphere,
which may be carried out by using the functional integral just given above.
For this purpose, we use the ``imaginary time trick'', that is,
we define $ {\tau} = - i z $, and introduce the ``wave function'' $ \Psi({\bf{S}}, \tau) $,
then we have the integral equation:
\begin{equation}
  \Psi({\bf S}, \tau) = \int K[{\bf S}, \tau \vert {\bf S}',0 ] \Psi({\bf S}', 0) d{\bf S}' . 
\end{equation}
where the propagator is written in terms of the ``imaginary time'' $ \tau $:
$$
K[{\bf S} ,\tau \vert {\bf S}', 0 ] = \int \exp \left[ \frac{i}{h}\int L_{S} d\tau \right] {\cal{D}}[{\bf{S}(\tau)}]  
$$
with the {\it Lagrangian} 
\begin{equation}
  L_{S} = \frac{ 1 }{ 2 } \left( \frac{ d{\bf S} }{ d \tau } \right)^2 + i \frac{d{\bf S}}{d\tau}\cdot {\bf A} - W 
\end{equation}
together with 
\begin{equation}
  W = \frac{ {\bf{A}}^{2} }{ 2 } - M h 
  \label{eq:ppote}
\end{equation}
Following the standard procedure of Feynman path integral \cite{Schulman}, we obtain the Schr{\"o}dinger-type equation:
\begin{eqnarray}
  i h \frac{ {\partial} {\Psi}}{ {\partial} {\tau} } = \frac{1}{2} \left( {\bf{p}} - i {\bf{A}} \right)^{2} \Psi + W \Psi
  \label{eq:schrotypeeq}
\end{eqnarray}
where ``momentum'' $   {\bf{p}} = - i h \frac{ {\partial} }{ {\partial} {\bf{S}} } \equiv  -i h \nabla $.
Here we note that the second term of Eq.~(\ref{eq:ppote}), which is
the contribution coming from the Jacobian, is crucial.
By using $ (\nabla \cdot {\bf A}) \Psi + {\bf A}\cdot \nabla \Psi = \nabla \cdot ({\bf A}\Psi) $ and
replacing the imaginary time $ {\tau} $ with the original real coordinate $ z $, namely $ {\tau} \rightarrow - i z $,
and getting ``wave function'' $ {\Psi} $ back to the original probability distribution $ P $,
we can derive the FP equation (alias Smolkowski equation):
\begin{eqnarray}
  \frac{ {\partial} P }{ {\partial} z } = \frac{ h }{ 2 } \left( \frac{ {\partial} }{ {\partial} {\bf{S}} } \right)^{2} P
                                             + \frac{ {\partial} }{ {\partial} {\bf{S}} } \cdot \left( {\bf{A}} P \right)
  \label{eq:fpeq}
\end{eqnarray} 
The FP equation obtained above  can be rewritten as the continuity equation:
\begin{equation}
  \frac{\partial P}{\partial z} + \nabla\cdot {\bf J} = 0 
\end{equation}
where $ {\bf J } $ denotes the probability current 
$ {\bf J } = - \left( \frac{h}{2}\nabla P + {\bf A}P \right) $,
which is written in terms of the components with respect to polar coordinate $ (\theta, \phi) $
\begin{equation}
  \begin{cases}
    J_{\theta}  =  -\frac{h}{2}\frac{\partial P}{\partial \theta} - \frac{P}{1+\mu^2} 
    \left(\mu\frac{\partial H}
      {\partial \theta} 
      -  \frac{1}{\sin\theta}\frac{\partial H}{\partial \phi}\right)  & \nonumber  \\
    J_{\phi} =    - \frac{h}{2} \frac{1}{\sin\theta}\frac{\partial P}{\partial \phi} - \frac{P}{1+\mu^2} 
    \left(\frac{\partial H}{\partial \theta} 
      + \frac{\mu}{\sin\theta}\frac{\partial H}{\partial \phi}\right) & \nonumber \\
  \end{cases}
\end{equation}
Thus the FP equation becomes
\begin{equation}
  \frac{\partial P}{\partial z} = - \frac{1}{\sin\theta} \left[ \frac{\partial}{\partial \theta}(\sin\theta J_{\theta}) 
    + \frac{\partial J_{\phi}}{\partial \phi} \right] .
\end{equation}

\section{The application to the optical rotation }
\label{sec:application}

We now apply  the general formulation presented in the previous section to the optical rotation.
This case is reduced to an essentially one-dimensional system, so it enables us to carry out analytic approach.
We will discuss the problem in two frameworks:
The one is to use the asymptotic limit of the functional integral and the other is based on the approximate scheme for the FP equation.

\subsection{Asymptotic analysis in the small diffusion limit}

The advantage of the functional integral is that
one can directly derive the transition probability from the one point to other on the Poincar{\'e} sphere.
As is well known in quantum mechanics, the evaluation of functional integral
can be achieved by adopting  approximate methods.
The simplest way is provided by an asymptotic method, which is realized in the small diffusion limit,
namely, $ h \sim 0 $, which corresponds to the semiclassical approximation in quantum mechanics.
In the limit of $ h \sim 0 $, the transition probability can be given by the stationary phase approximation.
Let the classical action function, $ S_{cl} $, be expressed as
\begin{eqnarray}
  S_{cl} = \int_{0}^{z} {\cal{L}}_{cl} dz = \int_{0}^{z} \left[ \frac{1}{2} \left( \frac{ d{\bf{S}} }{ dz } + {\bf{A}} \right)^2 \right] dz  \nonumber
\end{eqnarray}
where $ S_{cl} $ is evaluated by finding out the {\it classical orbit}
which is determined by the variational principle: $ \delta S_{cl} = 0 $.
Using this classical action, the transition probability is written as
\begin{equation}
 K \sim \exp \left[ - \frac{1}{h} S_{cl} \right]
\end{equation}
Here we note that the contribution of functional Jacobian disappears in the limit $ h \sim 0 $.

For the case of the optical rotation $ H = H(\theta) $,
the Lagrangian $ {\cal{L}}_{cl} $ is simply written in terms of the angle $ \theta $:
\begin{equation}
  {\cal{L}}_{cl} = \frac{ 1 }{ 2 } \left[ \left( \dot{\theta} + \frac{ \mu }{ 1 + {\mu}^2 } \frac{dH }{ d\theta } \right)^2 + {\sin}^2 \theta \left( \dot{\phi} 
  + \frac{ 1 }{ 1 + {\mu}^2 } \frac{ 1 }{ \sin \theta } \frac{dH }{d\theta } 
  \label{lag}
  \right)^2 \right]
\end{equation}
From this Lagrangian, we have the constant of motion:
\begin{equation}
  \sin^2\theta \left( \dot\phi + \frac{1}{1+ \mu^2} \frac{1}{\sin\theta}\frac{dH}{d\theta} \right)= {\cal{C}}
 \label{constant}
\end{equation}
Following the procedure of classical dynamics,
$ \phi $ is eliminated by using the {\it Routhe function}~\cite{LL_II}:
$ R = {\cal{C}} \dot\phi - {\cal{L}}_{cl}  $, which is given by
\begin{equation}
  R =  - \frac{1}{2} \left( \dot\theta +  \frac{ \mu }{1 + \mu^2}\frac{dH}{d\theta} \right)^2 
  + \frac{ {\cal{C}}^2 }{2\sin^2\theta} - \frac{ {\cal{C}}}{1+ \mu^2}\frac{1}{\sin\theta}\frac{dH}{d\theta}
  \label{Rouse}
\end{equation}
The equation of motion for $ \theta $ is derived by using this Routhe function,
$ \frac{ d }{ dt } \left( \frac{ \partial R }{ \partial \dot{\theta} } \right) - \frac{ \partial R }{ \partial \theta } = 0 $.
By substituting the solution of this equation into the Lagrangian $ {\cal{L}}_{cl} $,
we obtain the transition probability as the form
\begin{equation}
  K_{cl} = \exp \left[ - \int \frac{1}{2h} \left\{ \left( \dot\theta + \frac{ \mu }{1 + \mu^2} \frac{dH}{d\theta}  \right)^2
      + \frac{ {\cal{C}}^2 }{\sin^2\theta} \right\} dz \right] 
\end{equation}

The emergence of the constant of the motion $ \cal{C} $ is crucial,
which is a characteristics of the random fluctuation.
Here we adopt a perturbation scheme, namely, we treat $ C $ as a small parameter.
This procedure could be effective for concise evaluation for the transition $ K_{cl} $.
Hence it follows that the last two terms in Eq.~(\ref{Rouse}) may be omitted and
the equation of motion is simply given by the so called {\it instanton};
\begin{eqnarray}
   \dot\theta + \frac{ \mu}{1 + \mu^2}\frac{dH}{d\theta} = 0 
  \label{theta}
\end{eqnarray}
together with the equation of motion for the rotation angle, $ \phi $,
\begin{equation}
  \dot\phi +  \frac{1}{1+ \mu^2} \frac{1}{\sin\theta}\frac{dH}{d\theta}= 0  
  \label{phi}
\end{equation}
which is obtained by omitting the term $ \frac{ C }{ \sin^{2} \theta } $ in (\ref{constant}).
Two equations (\ref{theta}) and (\ref{phi}) are nothing but $ \theta $ and $ \phi $ component
of the equation  (\ref{eq:langevinnoise}) without the fluctuation $ \eta $.
By using this solution, we obtain 
\begin{equation}
  K_{cl} = \exp \left[ - \frac{ {\cal{C}}^2 }{ 2h } \int^z_{z_0} \frac{dz}{ \sin^2 \theta } \right] 
  = \exp \left[ - \frac{\mathcal{C}^2}{2h} f( \theta , {\theta}_0) \right]
\end{equation}
In what follows, we consider the linear and nonlinear optical rotation separately. \\

(I): The case of pure Faraday effect $ H = \gamma \cos\theta $, for which the equation of motion turns out to be
\begin{eqnarray}
  \frac{ d \theta }{ dz } = \frac{2}{\lambda} \frac{ \mu \gamma }{ 1 + {\mu}^{2} }  \sin{\theta}, 
  ~~~ \frac{ d \phi }{ dz } = \frac{2}{\lambda} \frac{ \gamma }{ 1 + {\mu}^{2} }
  \label{eq:spiral}
\end{eqnarray}
which gives the solution:
\begin{equation}
  \cos\theta = - \tanh \left[ \frac{2}{\lambda} \frac{\mu\gamma}{1+ \mu^2}z \right], \ \ 
  \phi = \frac{2}{\lambda} \frac{ \gamma }{ 1 + {\mu}^{2} } z 
  \label{spiral2}
\end{equation}

This solution shows a spiral structure; namely, at $ z\rightarrow -\infty $, $ \cos \theta \sim 1 $ (the left handed circular polarization),
whereas at $ z\rightarrow + \infty $, $ \cos \theta \sim - 1 $ (the right  handed circular polarization).
The angle $ \phi $ changes with a constant pitch.
This way, the orbit converges to the south pole as shown in Fig.2. 
%~\ref{fig:spiral}.
%
\begin{figure}[t]
  \begin{center}
   \includegraphics[width=60mm]{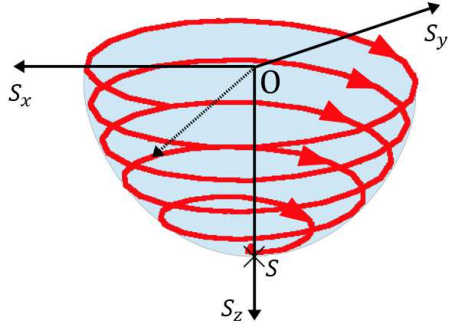}
    \caption{(Color Online).
      Orbit spiraling to the south pole on Poincar{\'e} sphere.}
   \label{fig:spiral}
  \end{center}
\end{figure}
The function $ f( \theta, \theta_{0} ) $ turns out to be the form
\begin{eqnarray}
  f(\theta, \theta_{0}) &=& \frac{ 1 + {\mu}^{2} }{ \mu \gamma } \int_{ {\theta}_{0} }^{\theta} \frac{ 1 }{ \sin^{3} \theta' } d \theta'  \nonumber \\
  &=& \frac{ 1 + {\mu}^{2} }{ 8 \mu \gamma } \left[ \tan^{2} \frac{ \theta }{ 2 } 
    + 4 \ln \left( \tan \frac{ \theta }{ 2 } \right) - \cot^{2} \frac{ \theta }{ 2 } \right]^{ \theta }_{ {\theta}_{0} }
\end{eqnarray}
If we put $ x = \tan \frac{ \theta }{ 2 } $ and $ x_{0} = \tan \frac{ {\theta}_{0} }{ 2 } $ (stereographic coodinate),
the transition probability is obtained as
\begin{eqnarray}
  \label{eq:trnamp_ia}
  K_{cl} = \left( \frac{ x }{ x_{0} } \right)^{ - \frac{ 2 {\cal{C}}^{2} }{ h } \frac{ 1 + {\mu}^{2} }{ 8 \mu \gamma } } 
  \exp \left[ - \frac{ {\cal{C}}^{2} }{ 2h } \frac{ 1 + {\mu}^{2} }{ 8 \mu \gamma } \left\{ \left( x^{2} - x_{0}^{2} \right) - \left( \frac{ 1 }{ x^{2} } - \frac{ 1 }{ x_{0}^{2} } \right) \right\} \right]
\end{eqnarray}

(II): Next we consider the pure nonlinear birefringence:
$ H = B \cos^{2} \theta $, for which we have
\begin{eqnarray}
  \frac{ d \theta }{ dz } = \frac{ \mu B}{ 1 + {\mu}^{2} }\sin{2 \theta}, 
  ~~~\frac{ d \phi }{ dz } = \frac{ 2B }{ 1 + {\mu}^{2} } \cos{\theta}
  \label{eq:nonbire}
\end{eqnarray}
Then we obtain the solution:
$ \cos 2 \theta = - \tanh \left[ \frac{ 2 \mu B }{ 1 + {\mu}^{2} } z \right] $, which leads to
\begin{eqnarray}
  \cos \theta = \pm \sqrt{ \frac{1}{2} \left( 1 - \tanh \left[ \frac{ 2 \mu B }{ 1 + {\mu}^{2} } z \right] \right) },
  ~~~  \phi = \frac{ 2B }{ 1 + {\mu}^{2} } \int \cos{\theta}(z) dz
\end{eqnarray}
This result indicates an asymptotic behavior of the Stokes parameters, namely,
in the infinite ``future'' $ z \rightarrow + \infty $, $ \cos \theta \sim 0 $, hence $ \dot{\phi} \sim 0 $.
It shows that we have the linear polarization for which the angle $\phi$ asymptotically tends to constant.
On the other hand, in the infinite ``past'' $ z \rightarrow - \infty $,
it follows that $ \cos \theta \sim \pm 1 $, so $ \dot{\phi} = {\rm constant} $,
which implies that the polarization starts with the circular polarization with the angle $\phi$ asymptotically proportional to $z$.
This feature is similar to the case of Faraday effect.
For this case, we have the function $ f(\theta, \theta_0) $ as follows:
\begin{eqnarray}
  f( \theta, \theta_{0}) = \frac{ 1 + {\mu}^{2} }{ 4 \mu B } \left[ 2 \ln \left( \tan \theta \right) - \cot^{2} \theta \right]^{ \theta }_{ {\theta}_{0} }
\end{eqnarray}
If we put $ y = \tan \theta $ and $ y_{0} = \tan {\theta}_{0} $,
we obtain the transition probability:
\begin{eqnarray}
  \label{eq:trnamp_ib}
  K_{cl} = \left( \frac{ y }{ y_{0} } \right)^{ - \frac{ {\cal{C}}^{2} }{ h } \frac{ 1 + {\mu}^{2} }{ 4 \mu B } } 
  \exp \left[ - \frac{ {\cal{C}}^{2} }{ 2h } \frac{ 1 + {\mu}^{2} }{ 4 \mu B } \left( \frac{ 1 }{ y_{0}^{2} } - \frac{ 1 }{ y^{2} } \right) \right]
\end{eqnarray}

{\it Discussion of $ K_{cl} $}: 
Note that for both cases (I) and (II), the functions $ f(\theta, \theta_0) $ vanishes at
$ \theta = \pi  ( x = \infty) $, (right handed circular polarization),
and $ \theta = \frac{\pi}{2} ( y = \infty) $ (linear polarization) respectively.
This vanishment may come from the nature of the orbit on the Poincare sphere:
Namely, the orbit concentrates to the pole with uniform rotation about the $ z $ axis.
This suggests that the orbit never reaches these two limiting orbits.
The feature of the concentration of orbit may be considered to be natural consequence of a presence of the dissipation.

{\it Remark}: 
Using the instanton solution assisted by the constant of the motion $ \cal{C} $,
one can obtain a concise expression for the transition probability.
Without this procedure, it is inevitable to adopt more complicated way to deduce meaningful physical results.
However, the use of instanton is effective only when we adopt the {\it perturbative} assumption that the constant of motion
$ {\cal{C}} $ is small enough, which would be comparable order with the square root of the diffusion constant.

\subsection{Analysis of the FP equation}
\label{sec:analysis_fp}

The asymptotic analysis given above is restricted to a specific aspect of the behavior of Stokes parameters in the small diffusion limit,
so it is inevitable to analyze the FP equation itself in order to scrutinize the details of the interplay between the dissipation effect and diffusion.

In order to carry out this analysis, we first consider the stationary equation $ \frac{\partial P}{\partial z} = 0 $,
and put an ansatz of the Boltzmann distribution:
$ P({\bf S}) = \exp[-\beta H] $ with $ \beta $ the inverse of the {\it effective temperature} $ \beta = \frac{1}{T} $
(which is not real temperature in statistical thermodynamical sense).
By substituting this into the right hand side of the FP equation, we obtain
\begin{equation}
  \left( \beta h - \frac{2 \mu}{1 + \mu^2} \right) \left[ \nabla^2 H - \beta(\nabla H)^2 \right] = 0.  \nonumber 
\end{equation}
This equation is satisfied for arbitrary $H$, only if the following condition holds:
\begin{equation}
  \beta h - \frac{2 \mu}{1 + \mu^2} = 0. 
  \label{FD}
\end{equation}
This is the well known fluctuation--dissipation theorem, which establishes the relation
between the dissipation coefficient $ \mu $ and the diffusion (fluctuation) constant $ h $.
The Boltzmann distribution is a consequence of the balance between dissipation and diffusion;
namely, without dissipation one cannot have a stable equilibrium state in {\it thermodynamical sense}.

{\it The diffusion behavior on the Poincar{\'e} sphere}:
Now we shall analyze the FP equation in the non--equilibrium state.
For the present optical rotation, in which the Hamiltonian $H$ does not depend on the coordinate $\phi$,
it is natural to assume that $P$ is independent of $\phi$.
Then we have the FP equation in the following form:
\begin{eqnarray}
  \label{eq:fpeqeq}
  \frac{ \partial P }{ \partial z } = \frac{1}{\sin \theta} \frac{ \partial }{ \partial \theta } \left[ \sin \theta \left\{ \frac{h}{2} \frac{ \partial P }{ \partial \theta } 
  + \frac{\mu}{1 + {\mu}^{2} } \left( \frac{ \partial H }{ \partial \theta } P \right) \right\}  \right]
\end{eqnarray}
Here putting $ P = e^{- \varepsilon z } f(\theta)$, we obtain the {\it eigenvalue equation}~\cite{Brown}
\begin{eqnarray}
  \label{eq:fpreplace}
  - \varepsilon f(\theta) = \frac{1}{\sin \theta} \frac{d}{d \theta} \left[ \sin \theta \left\{ \frac{h}{2} 
  \frac{d f(\theta)}{d \theta} + \frac{\mu}{1 + {\mu}^{2}} \left( \frac{ dH }{ d \theta } f(\theta) \right)  \right\}  \right]
 \label{eigen}
\end{eqnarray}
By noting that Eq.~(\ref{eq:fpreplace}) includes the parameter $\mu$,
one may think of carrying out the perturbation scheme if $\mu$ is small ($\vert \mu \vert \ll 1$).
That is, one starts with the equation for $\mu = 0$ as an unperturbed solution, which is given by the Legendre polynomial.
However this procedure may not be relevant, since the case $\mu = 0$ corresponds to $\beta = 0$,
which does not represent equilibrium state in the thermodynamical sense, as it is shown by Eq.~(\ref{FD}).
From this inspection, it is suitable to consider the case in which $\vert \mu \vert$ is not small,
$ \vert \mu \vert > 1 $, which we call the ``strong coupling scheme''.

Let us now consider the procedure of eigenvalue problem:
Here by putting $x = \cos \theta$, the eigenvalue equation (\ref{eigen}) can be rewritten as
\begin{eqnarray}
  \frac{d}{ dx } \left[ ( 1 - x^{2} ) e^{ - \beta H } \frac{ d }{ dx } \left( e^{\beta H} f \right) \right] + \frac{ 2 \varepsilon }{ h } f = 0 
\end{eqnarray}
where $\beta = \frac{2 \mu}{h (1+{\mu}^{2})}$.
Furthermore, we assume $ f = e^{ - \beta H } g $, then
\begin{eqnarray}
  \label{eq:eigenvalue}
  \frac{d}{ dx } \left[ ( 1 - x^{2} ) e^{ - \beta H } \frac{ dg }{ dx } \right] = - \frac{ 2 \varepsilon }{h} e^{ - \beta H } g
\end{eqnarray}
Let us define the functionals $D$ and $N $ as
\begin{eqnarray}
  \label{eq:dd}
  D = \int^1_{-1} ( 1 - x^{2} ) e^{ - \beta H } \left( \frac{ dg }{ dx } \right)^{2} dx ~~,~~
  \label{eq:hh}
  N = \int^1_{-1} e^{ - \beta H } g^{2} dx
\end{eqnarray}
By using Eq.~(\ref{eq:eigenvalue}) and Eq.~(\ref{eq:dd}), we have the following relation
\begin{eqnarray}
  \frac{ 2 \varepsilon }{ h } = \frac{ D }{ N }
\end{eqnarray}
In addition, we set the orthogonality relation between two eigenstates $g$ and ${\tilde{g}}$
\begin{eqnarray}
  \label{eq:00}
  \int^1_{-1} g {\tilde{g}} e^{ - \beta H } dx = 0.
\end{eqnarray}
In this way, the problem is reduced to find (local) minima of $D$ with the constraint $ N  = 1 $.
As a special case, we have the lowest eigenvalue $\varepsilon = 0 $, for which we see that $g$ becomes $constant$.
Namely, the function $f$ has the form of Boltzmann distribution $f \propto e^{ - \beta H }$.

We can thus construct a sequence of the eigenfunctions; $ \{g_n(x)~(n=1, 2, \cdots \} $.
Here to be noted is that one does not need to consider higher order terms of $g_{n} (x)$ ($n \geq 2$),
because they may not give a significant contribution as will be explained shortly.
Now let us try to find the first ``excited'' state in an explicit form.
We put
\begin{eqnarray}
  \label{eq:firstexcited}
  g_{1} (x) = a + b x + c x^{2}
\end{eqnarray}
The coefficients $a$, $b$ and $c$ are determined by three conditions corresponding to Eqs.~(\ref{eq:dd}) and (\ref{eq:00}),
which are written explicitly as
\begin{eqnarray}
  \label{eq:ii}
  D = J_{0} b^{2} + 4 J_{1} b c + 4 J_{2} c^{2} \equiv \epsilon_{1}
\end{eqnarray}
and
\begin{eqnarray}
  \label{eq:constrainta}
  a^{2} K_{0} + 2ab K_{1} + \left( b^{2} + 2ca \right) K_{2} + 2bc K_{3} + c^{2} K_{4} = 1 \nonumber \\
  a K_{0} + b K_{1} + c K_{2} = 0
\end{eqnarray}
Here we put $ J_{j} = \int_{-1}^{1} ( 1 - x^{2} ) x^{j} e^{ - \beta H } dx $ and
$ K_{j} = \int_{-1}^{1} x^{j} e^{ - \beta H } dx $.
The minimum value of $D$ is obtained as follows:
First, by eliminating $a$ from Eq.~(\ref{eq:constrainta}),
one gets the constraint with respect to $b$, $c$:
\begin{eqnarray}
  \label{eq:cob}
  G(b, c) = \left( K_{2} - \frac{ K_{1}^{2} }{ K_{0} } \right) b^{2} + 2 \left( K_{3} - \frac{ K_{1} K_{2} }{ K_{0} } \right) bc 
  + \left( K_{4} - \frac{ K_{2}^{2} }{ K_{0} } \right) c^{2} - 1 = 0
\end{eqnarray}
Next, by using this constraint and following the Lagrange multiplier method,
we have the relations
\begin{eqnarray}
  \frac{ \partial }{ \partial {\chi}_{i} } ( D - \lambda G ) = 0 \ \ \ ( {\chi}_{i} = b, c, \lambda)
\end{eqnarray}
where ${\lambda}$ is a multiplier.
By solving this, we can obtain the value of $\epsilon_{1}$.

Having obtained $ \epsilon_1 $ (and hence the function $ g_1(x) $), the distribution function is constructed by
expanding in terms of $ g_0, ~g_1 $. That is, $P (x,z)$ can be written as
\begin{eqnarray}
  \label{eq:excitedp}
  P(x,z) = \{ C_0 g_0 (x) + C_1\exp[-\epsilon_1 z]g_1(x) \} \exp \left[ - \beta H \right].
\end{eqnarray}
The coefficients $C_{0} $ and $C_{1} $ are determined by the initial conditions, which are given as follows:
$ C_n = \int^1_{-1}P (x,0) g_{n} (x) \exp \left[ - \beta H \right] dx $.
The distribution of this form may be utilized to calculate the time evolution of
the average value of physical quantities under consideration.
One can derive the relaxation length for the expectation value,
which may be written in a form $ \langle G \rangle = G_0 + G_1\exp[-\epsilon_1 z] $, showing the relaxation behavior.
The relaxation length is thus given by $ l \sim 1/\epsilon_1 $.
For $ z \gg l $, the distribution becomes the stationary value.

{\it Alternative procedure}:
The above procedure to evaluate $ \epsilon_1 $ looks rather complicated.
So it is desirable to look for an alternative way to guess the
distribution function without recourse to derivation of $ \epsilon_1 $.
To carry out this we adopt the following scheme:
First we choose a physical quantity $ G(\theta ) $, which is a function of $ \cos\theta $,
and consider the expectation value: $ \langle G(\theta ) \rangle = \int G P d\Omega $.
The next step is to find the form of the distribution function by the method of a trial function.
Being inferred from the fact that $ P(x,z) $ is written as the quadratic function of $ x (\equiv \cos\theta) $,
we give $ P(x,z) $ by using the Hamiltonian $ H ( \theta ) $ in the following form:
$ P = \left[ A_{0} + A_{1} H ( \theta ) F (z) \right] e^{ - \beta H(\theta)} $
with the coefficients $ A_0 $, $ A_1 $ being appropriately chosen.
$ F(z) $ is a trial function that should be determined by the evolution
equation for the expectation value of $ G ( \theta ) $.
By using FP equation, we obtain
\begin{eqnarray}
  \label{eq:fp_integral}
  \int G (\theta) \frac{ \partial P }{ \partial z } d \Omega
  &=& \frac{ h }{ 2 } \int G (\theta) \nabla^{2} P d \Omega + \int G( \theta ) \nabla \cdot \left( {\bf{A}} P \right) d \Omega  \nonumber \\
  &=& \frac{h}{2} \int P \left( \nabla^{2} G \right) d \Omega - \int \left( \nabla G \right) \cdot \left( {\bf{A}} P \right) d \Omega
\end{eqnarray}
Here it is natural to choose  $ G ( \theta ) = \cos \theta $, that is, the ellipticity, hence
Eq.~(\ref{eq:fp_integral}) can be rewritten as the differential equation for $ F(z) $:
\begin{eqnarray}
  K \frac{ d F(z) }{ dz } = L + M F(z)
 \label{relaxation} 
\end{eqnarray}
where
\begin{eqnarray}
  K &=& A_{1} \int \cos \theta H( \theta ) e^{ - \beta H(\theta) } d \Omega  \nonumber \\
  L &=& - h A_{0} \int \cos \theta e^{ - \beta H(\theta) } d \Omega
        + \frac{ \mu }{ 1 + \mu^{2} } A_{0} \int \sin \theta \frac{ dH }{ d\theta } e^{ - \beta H(\theta) } d \Omega \nonumber  \\
  M &=& - h A_{1} \int \cos \theta H( \theta ) e^{ - \beta H(\theta) } d \Omega 
        + \frac{ \mu }{ 1 + \mu^{2} } A_{1} \int \sin \theta H( \theta ) \frac{ dH }{ d\theta } e^{ - \beta H(\theta) } d \Omega 
\end{eqnarray}
Equation (\ref{relaxation}) is solved as 
$ F(z) = - \frac{ L }{ M } \left( 1 - \exp \left[ \frac{ M }{ K } z \right] \right) $
from which we see that the value $  \frac{ M }{ K } $ corresponds to $ - \epsilon_1 $.
The  concrete form for $F(z)$ is crudely obtained as follows:
By using  $ H = \gamma \cos \theta + B \cos^{2} \theta $ and 
considering the $ \beta \rightarrow 0 $ (high temperature limit, which means $ \exp[-\beta H] \sim 1$),
$ K $, $ L $ and $ M $ are evaluated as follows:
\begin{eqnarray}
  K &=& \frac{4}{3} \pi A_{1} \gamma  \nonumber \\
  L &=& - \frac{8}{5} \frac{ \mu \pi }{ 1 + \mu^{2} } A_{0} B \gamma \nonumber \\
  M &=& - \frac{4}{3} \pi h A_{1} \gamma - \frac{8}{5} \frac{ \mu \pi }{ 1 + \mu^{2} } A_{1} B \gamma
\end{eqnarray}
After sufficiently long time, it follows that $ F(\infty) = - L/M $, which means that the distribution function turns out to be
\begin{equation}
  P(x, z) \sim \left( A_0 - \frac{ A_1 L }{ M } H \right) \exp \left[ - \beta H \right]. 
  \label{distribution}
\end{equation}
This form deviates from the usual Boltzmann distribution.
However, it  can be reduced to the modified Boltzmann equation:
Namely, owing to the fluctuation dissipation theorem; $ \frac{ \mu }{ 1 + \mu^{2} } = (1/2) \beta h $,
we see that $ A_{1} L/M $ is proportional to $ A_{0} \beta $.
As far as we are concerned with the high temperature limit $ \beta \rightarrow 0 $,
the distribution tends to the modified Botzmann distribution
which is described by the renormalized inverse temperature $ \beta' = \beta \left( 1 + \frac{1}{ \frac{5}{3B} + \beta } \right) $.

Thus by evaluating the function $ F(z) $, one could  avoid a complicated procedure for calculating $ \epsilon_1 $,
though we have to make sacrifice for that there is a discrepancy from the correct Boltzmann distribution.

\section{Summary}
\label{sec:summary}

On the basis of the soliton solution for two component nonlinear Schr{\"o}dinger equation,
a novel stochastic theory has been presented for the  polarization evolution.
By taking into account the effect of the dissipation and the randomness inherent in the birefringent media,
the Langevin equation is derived for the Brownian motion of the Stokes parameters,
which is converted to the Fokker-Planck (FP) equation by using the functional integral.

In particular, we have given the analysis for the optical rotation.
The behavior of the Stokes parameters are governed by the ellipticity,
which enables one to bring a wealth of experimental informations about the polarized light.
Indeed, in Section~\ref{sec:application},
we have discussed the cooperative effects for the randomness and dissipation inherent in anisotropic media,
which was investigated from two aspects:
The asymptotic limit for path integral and the diffusion analysis of the FP equation.
These consequences would provide with a clue to the stochastic approach which would be launched future.
The present approach is, however, still limited in the sense that the optical gyration,
for which the Hamiltonian has $ \phi $ dependence, could not be treated properly.
This case may require a new method and is left open to future study.

\begin{appendix}

\section{Derivation of the birefringence }
\label{append:derivation}

\subsection{Nonlinear birefringence $V_{\rm{NL}}$  }

In order to derive $ V_{\rm{NL}}$, we start with the displacement field of the third order; that is written as ${\bf{D}}^{(3)}$~\cite{LL}
\begin{eqnarray}
  {\bf{D}}^{(3)} = g_{0} \vert {\bf{E}} \vert^{2} {\bf{E}} - g \left( E_{1}^{2} + E_{2}^{2} \right) {\bf{E}}^{*}
\end{eqnarray}
where ${\bf{E}} = {\bf{f}} e^{ i k n_{0} z }$, so we have
\begin{eqnarray}
  \label{eq:d3}
  {\bf{D}}^{(3)} = g_{0} \vert {\bf{f}} \vert^{2} {\bf{f}} e^{ i k n_{0} z } - g \left( f_{1}^{2} + f_{2}^{2} \right) {\bf{f}} e^{ i k n_{0} z } = \hat{v}_{\rm{NL}} {\bf{f}} e^{ i k n_{0} z }
\end{eqnarray}
The second term is expressed as
\begin{eqnarray}
  g \left( E_{1}^{2} + E_{2}^{2} \right) {\bf{E}}^{*} &=& g
  \left(
    \begin{array}{cc}
      E_{1}^{*} E_{1} & E_{1}^{*} E_{2} \\
      E_{2}^{*} E_{1} & E_{2}^{*} E_{2}
    \end{array} 
  \right)
  \left(
    \begin{array}{c}
      E_{1} \\
      E_{2}
    \end{array} 
  \right)   \nonumber \\
  &=& g
  \left(
    \begin{array}{cc}
      f_{1}^{*} f_{1} & f_{1}^{*} f_{2} \\
      f_{2}^{*} f_{1} & f_{2}^{*} f_{2}
    \end{array} 
  \right)
  \left(
    \begin{array}{c}
      f_{1}  \\
      f_{2} 
    \end{array} 
  \right)
    \nonumber \\
  &=& \left[ g
  \left(
    \begin{array}{cc}
      \frac{ \vert f_{1} \vert^{2} + \vert f_{2} \vert^{2} }{ 2 } & 0 \\
      0 & \frac{ \vert f_{1} \vert^{2} + \vert f_{2} \vert^{2} }{ 2 }
    \end{array} 
  \right)
  + g
  \left(
    \begin{array}{cc}
      \frac{ \vert f_{1} \vert^{2} - \vert f_{2} \vert^{2} }{ 2 } & f_{1}^{*} f_{2} \\
      f_{2}^{*} f_{1} & - \frac{ \vert f_{1} \vert^{2} - \vert f_{2} \vert^{2} }{ 2 }
    \end{array} 
  \right)
\right]
  \left(
    \begin{array}{c}
      f_{1}   \\
      f_{2} 
    \end{array} 
  \right)
\end{eqnarray}
where we omit the plane wave factor $ e^{ i k n_{0} z} $ in the last two lines.
Thus, $ \hat{v}_{\rm{NL}} $ in Eq.~(\ref{eq:d3}) becomes
\begin{eqnarray}
  \hat{v}_{\rm{NL}} = \frac{ G_{0} }{2} \vert f \vert^{2} {\bf{1}} - g
  \left(
    \begin{array}{cc}
      \frac{ \vert f_{1} \vert^{2} - \vert f_{2} \vert^{2} }{ 2 } & f_{1}^{*} f_{2} \\
      f_{2}^{*} f_{1} & - \frac{ \vert f_{1} \vert^{2} - \vert f_{2} \vert^{2} }{ 2 }
    \end{array} 
  \right)
\end{eqnarray}
where $ \frac{G_{0}}{2} = g_{0} + \frac{g}{2} $.
Using the component in terms of the circular polarization; $ (\psi_1, \psi_2) $, it follows that 
\begin{eqnarray}
  f_1^{*}f_1 - f_2^{*}f_2 & = & \psi_1^{*}\psi_2 + \psi_2^{*}\psi_1  \nonumber \\
  f_1^{*}f_2 & = & - i (\vert {\psi}_{1} \vert^{2} - \vert {\psi}_{2} \vert^{2} - {\psi}_{1}^{*} {\psi}_{2} + {\psi}_{2}^{*} {\psi}_{1}) \nonumber \\
  f_2^{*}f_1 & = & i (\vert {\psi}_{1} \vert^{2} - \vert {\psi}_{2} \vert^{2} - {\psi}_{2}^{*} {\psi}_{1} + {\psi}_{1}^{*} {\psi}_{2} )
\end{eqnarray}

Furthermore using the transformation for the Pauli spin:
\begin{equation}
  \hat U\sigma_1 \hat U^{-1} = \sigma_2,  ~~\hat U\sigma_2 \hat U^{-1} = \sigma_3,  ~~\hat U\sigma_3 \hat U^{-1} = \sigma_1
\end{equation}
we have the expression: 
\begin{eqnarray}
  V_{\rm{NL}} = T \hat{v}_{\rm{NL}} T^{-1} = \frac{ G_{0} }{2} \vert \psi \vert^{2} {\bf{1}} - g
  \left(
    \begin{array}{cc}
      - \frac{ \vert {\psi}_{1} \vert^{2} - \vert {\psi}_{2} \vert^{2} }{ 2 } & {\psi}_{2}^{*} {\psi}_{1} \\
      {\psi}_{1}^{*} {\psi}_{2} & \frac{ \vert {\psi}_{1} \vert^{2} - \vert {\psi}_{2} \vert^{2} }{ 2 }
    \end{array} 
  \right)
\end{eqnarray}

\subsection{Linear birefringence $ V_{\rm{L}} $}

We derive the potential term for linear birefringence that arises from the
external electro-magnetic effect, which consists of the Faraday and Kerr effects, so
we write $ \hat{v}_{\rm{L}} = \hat{v}_{\rm{L}}^{\rm{F}} + \hat{v}_{\rm{L}}^{\rm{K}} $.
First the Faraday effect is given by the matrix
\begin{eqnarray}
  \hat{v}_{L}^{\rm{F}} = 
  \left(
    \begin{array}{cc}
    0& i\gamma  \\
      -i\gamma  & 0 
    \end{array} 
  \right)
\end{eqnarray}
with $ \gamma = C {\rm{H}}_{\rm{ex}} $, where $ {\rm{H}}_{\rm{ex}}$ means the external magnetic field, and $ C $ is called the Verde constant.
Next we consider the Kerr effect:
By putting $e_{1}$ and $e_{2}$ as the external electric fields
which represent the component perpendicular to the direction of the wave propagation,
$ \hat{v}_{\rm{L}}^{\rm{K}} $ is given as the matrix
\begin{eqnarray}
  \hat{v}_{\rm{L}}^{\rm{K}} = G
  \left(
    \begin{array}{cc}
      e_{1}^{2} & e_{1} {e}_{2} \\
      e_{2} e_{1} & e_{2}^{2}
    \end{array} 
  \right)
  = \frac{G}{2} \left( e_{1}^{2} + e_{2}^{2} \right) {\bf{1}} + G
  \left(
    \begin{array}{cc}
      \frac{ e_{1}^{2} - e_{2}^{2} }{ 2 } & e_{1} e_{2} \\
      e_{2} e_{1} & - \frac{ e_{1}^{2} - e_{2}^{2} }{ 2 }
    \end{array} 
  \right)
\end{eqnarray}
Here, we introduce the constants $\alpha$, $\beta$ and $\gamma$ as
\begin{eqnarray}
  \hat{v}_{\rm{L}} = \hat{v}_{\rm{L}}^{\rm{F}} + \hat{v}_{\rm{L}}^{\rm{K}} \equiv \frac{G}{2} \left( e_{1}^{2} + e_{2}^{2} \right) {\bf{1}} +
  \left(
    \begin{array}{cc}
      \alpha & - \beta + i \gamma \\
      - \beta - i \gamma & - \alpha
    \end{array} 
  \right)
\end{eqnarray}
Then, writing it in terms of the circular polarization basis, we obtain
\begin{eqnarray}
  V_{\rm{L}} = U \hat{v}_{\rm{L}} U^{-1} = \frac{G}{2} \left( e_{1}^{2} + e_{2}^{2} \right) {\bf{1}} +
  \left(
    \begin{array}{cc}
      \gamma & \alpha - i \beta \\
      \alpha + i \beta & - \gamma
    \end{array} 
  \right)
\end{eqnarray}

\section{Gyration motion of the Stokes parameters}
\label{append:gyration}

We take up the birefringence arranged such that $ {\alpha} \neq 0 ~,~ {\beta} = {\gamma} = 0 ~,~ B \neq 0 $,
which describes the simultaneous effect of external Kerr effect and the nonlinear birefringence.
This case may be called the {\it optical gyration}.
For this case, the equation of motion for the Stokes parameters is written as
\begin{equation}
  \label{eq:dsdz}
  \frac{dS_x}{dz} = - 2 B S_{y} S_{z}, 
  ~\frac{dS_y}{dz} =
    - \left( \frac{2}{\lambda} \alpha - 2 B S_{x} \right) S_{z}, 
    ~ \frac{dS_z}{dz} = \frac{2}{\lambda} \alpha S_{y} 
\end{equation}
Here, we have two constants of motion: namely,
$ \frac{2 \alpha}{\lambda} S_{x} + B S_{z}^{2} = E $, which is just the Hamiltonian,
together with the magnitude of the Stokes parameters, $S_{x}^{2} + S_{y}^{2} + S_{z}^{2} = 1 $.
By choosing $ E = B $, we can obtain the single equation for the third component of the Stokes parameters,
which is written in terms of the elliptic integral:
\begin{eqnarray}
  \label{eq:ellipt}
  \int \frac{ d S_{z} }{ \sqrt{ \left( 1 - S_{z}^{2} \right) \big(1 - {\kappa}^{2} + {\kappa}^{2} S_{z}^{2} \big) } } = \frac{ 2 \alpha }{ \lambda } z
\end{eqnarray}
where we put $ \kappa = \frac{ \lambda B }{ 2 \alpha } $.
According to the value of $ \kappa $, $S_{z}$ is given by two kinds of Jacobi's elliptic functions~\cite{Toda2001}.
\begin{eqnarray}
  \label{eq:elliptsolu}
  S_{z} =
  \begin{cases}
    {\rm{cn}} \left( \frac{ 2 \alpha }{ \lambda } z, \kappa \right) \ \ {\rm{for}} \ \ \vert \kappa \vert < 1  \\
    {\rm{dn}} \left( \frac{ 2 \alpha }{ \lambda } \kappa z, \frac{1}{\kappa} \right) \ \ {\rm{for}} \ \ \vert \kappa \vert > 1
  \end{cases}
\end{eqnarray}
For the critical case, $ \vert \kappa \vert = 1 $, we have
$ S_z = \left( \cosh \left[ \frac{ 2 \alpha }{ \lambda } z \right] \right)^{-1} $.
By using Eq.~(\ref{eq:dsdz}), the remaining components ($S_{x}$, $S_{y}$) are obtained by using the equation of motion, which will be omitted here.

\section{Comment on the equilibrium state}

In what follows we address a brief description for the characteristic feature of the Boltzmann distribution.
The quantity we are concerned with, that is to be observed in actual experiment,
is the average value of the ellipticity with the equilibrium state.
This can be obtained by calculating the partition function (free energy):
\begin{equation}
  Z = \int e^{ - \beta H } d\Omega 
  (d\Omega= \sin\theta d\theta d\phi ) 
\end{equation}
and $ F = - \frac{1}{\beta} \log Z $.
For the case that both the linear and nonlinear optical rotation coexist, we have
\begin{eqnarray}
  Z = 2 \pi \exp \left[ \frac{ \beta {\gamma}^{2} }{ 4 B } \right] \int_{-1}^{1} \exp \left[ - \beta B \left( x + \frac{ \gamma }{ 2 B } \right)^{2} \right] dx
\end{eqnarray}
Here, we take limit $ \sqrt{ \beta B } \rightarrow \infty $, it follows that
\begin{equation}
  Z \sim \int^1_{-1} \exp[ -\beta(Bx^2 + \gamma x)] dx 
  = 2 \pi \sqrt{\frac{ \pi }{ \beta B } } \exp \left[ \frac{ \beta \gamma^2}{ 4 B} \right]
\end{equation}
Furthermore, if we put $ \gamma = 0 $, it turns out to be $  Z \rightarrow 2 \pi \sqrt{ \frac{ \pi }{ \beta B } } $.
For the linear rotation $ H = \gamma \cos \theta $, $ Z $ is not obtained in the limit of $ B = 0 $, so that
it should be evaluated in a direct way, which leads to $ Z = \frac{4\pi}{\beta\gamma} \sinh[\beta\gamma] $.
Thus the average value for the ellipticity can be calculated by using the relation
\begin{equation}
  \langle \cos\theta \rangle = \frac{\partial F}{\partial \gamma} 
\end{equation}
By using the free energy obtained above, for the linear rotation we have
\begin{equation}
  \langle \cos\theta \rangle = \frac{ 1 }{ \beta \gamma } - \frac{ 1 }{ \tanh \left( \beta \gamma \right) }
\end{equation}
whereas for the case of the coexistence of the linear and nonlinear birefringence;
we obtain $ \langle \cos \theta \rangle = - \frac{ \gamma }{ 2 B } $, which means that
for the pure nonlinear birefringence; $ \gamma = 0 $, we obtain $ \langle \cos \theta \rangle = 0 $.
This implies that the statistical average of the ellipticity can be observed only for the Faraday effect. \\

\end{appendix}

%\end{paper}
%--------------------------

\end{document}